# Considerations on time resolution of neutron irradited single pixel 3D structures at fuences up to $10^{17}$ $n_{eq}/cm^2$ using 120 GeV SPS pion beams

Evangelos-Leonidas Gkougkousis[a], Edgar Lemos Cid[b], Viktor Coco[b]

[a] University of Zurich
[b] European Organization for Nuclear Research (CERN), Geneva, Switzerland

*Abstract*

The proven radiation hardness of silicon 3D devices up to fluences of $1 \times 10^{17}$ $n_{eq}/cm^2$ makes them an excellent choice for next generation trackers, providing < 10 μm position resolution at a high multiplicity environment. The anticipated pile-up increase at HL-LHC conditions and beyond, requires the addition of < 50 ps per hit timing information to successfully resolve displaced and primary vertices. In this study, the timing performance, uniformity and efficiency of neutron and proton irradiated single pixel 3D devices is discussed. Fluences up to $1 \times 10^{17}$ $n_{eq}/cm^2$ in three different geometrical implementations are evaluated using 120 GeV SPS pion beams. A MIMOSA-26 type telescope is used to provide detailed tracking information with a ~ 5 μm position resolution. Productions with single- and double-sided processes, yielding active thicknesses of 130 and 230 μm respectively, are examined with varied pixel sizes from $55 \times 55$ μm$^2$ to $25 \times 100$ μm$^2$ and a comparative study of field uniformity is presented with respect to electrode geometry. The question of electronics bandwidth is extensively addressed with respect to achievable time resolution, efficiency and collected charge, forming a 3D phase space to which an appropriate operating point can be selected depending on the application requirements.

1. **Introduction**

Over the past decade, 3D pixel sensors have emerged as tracking devices in high-energy physics experiments, within environments of intense radiation fluxes, encountered in the |η| < 2 regions of ATLAS Insertable B-Layer [1] and ATLAS Forward Proton (AFP) detector [2]. The sensor design, decoupling charge-generating volume from the drift distance, accommodates shorter electrode spacing, consequently decreasing the charge carrier trapping probability. Recent studies using the Transient Current Technique (TCT) [3, 4], have further substantiated the exceptional timing performance of such devices. For vertically incident events, the orthogonal relationship between drift direction and particle trajectory results in an absence of Landau fluctuations, key factor in extending theoretical timing performance of such devices, primarily constrained only by the signal's drift time.

This excellent performance is nevertheless impacted by field non-uniformities, intrinsic to the column geometry of the collection electrodes. The resulting radially expanding field within the pixel volume increases signal time-jitter, degrading time resolution. Such a geometry, though detrimental under normal operation, can lead to high field densities near the collection electrode (> 30 V/μm) at higher bias voltages (> 500 V) typically used after irradiation. In this operating mode, impact ionization and charge multiplication occur near the collection electrode in a similar way as in Multi-Wire Proportional Chambers (MWPCs) and can compensate for trapping induced charge collection efficiency issues.

In this study, three different geometries are examined after proton and neutron irradiation, using 120 GeV SPS pion beams [5]. Pixel sizes of 55 x 55 μm$^2$ [6], 25 x 100 μm$^2$ and 50 x 50 μm$^2$ [7] in substrates of 230 μm (190 μm active depth) for the first geometry and 150 μm (130 μm active depth) for the second and third, are tested to establish the minimum active thickness and pixel size still yielding sufficient charge and low enough jitter to achieve a 30 ps time resolution. Questions of signal integrity, bandwidth and efficiency are treated.

2. **Timing and Signal Integrity**

Studies under a $^{90}$Sr source of the 50 x 50 μm variety single pixel structure (Figure 1 top), yield collected charge of $1.73 \pm 0.02$ fq (10,830 e$^-$), in agreement with the theoretically predicted value of ~ 82 e/μm for a MIP in fully depleted Si [8] at a bias voltage of 20 V at -20 °C. Signals are characterized by a fast rise time (10 % - 90 %) of $175 \pm 4$ ps, but present a tail at the slower end, characteristic of filed non-uniformities. Although charge per micrometer in the 3D device is reduced by a factor of 60 with respect to a typical LGAD[1] ($34.3 \pm 0.4$ fq for 50 μm thick HPK[2] device), an almost factor of 6 gain in rise time is noted ($980 \pm 1.4$ ps for LGADs - Figure 1 bottom).

Analog time resolution of the LGAD-3D system can be approximated as the sum of a Landau fluctuations term, field related distortion effects and the noise-induced jitter [9]. Through a Constant Fraction Discriminator (CFD) time walk correction approach, a 2D time resolution map is established with respect to the CFD of each of the components (Figure 2). For LGADs, selecting higher CFDs to profit from increased signal to noise ratio incurs a penalty in slew rate, deteriorating time resolution, thus resulting in the characteristic S curve along the x-axis. The plateau at ~ 40 % of peak amplitude corresponds to a time when all primary charges have reached the gain layer. In the case of 3Ds, due to the decoupling of drift and charge generation volumes and reduced electrode distances, the slew rate remains practically constant. Absence of gain under normal operation can be partially addressed by selecting higher CFDs, without degradation on time resolution.

---

[1] *Low Gain Avalanche Diode.*
[2] *Hamamatsu Photonics LTd.*

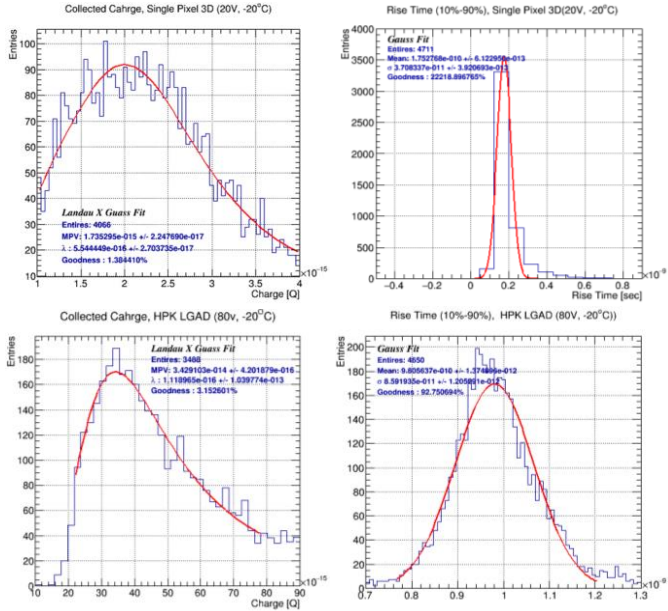

Fig. 1. Charge and rise time disitributions of a 150 $\mu m$, 50 x 50 $\mu m^2$ pitch, 3D single pixel devnice (top) and a 50 $\mu m$ thick LGAD (bottom).

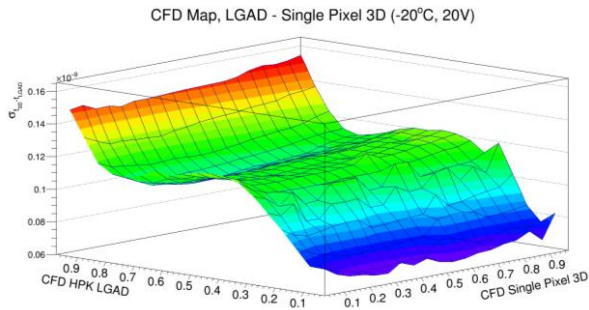

Fig. 2. Time resolution map of an LGAD - 3D (50 x 50 $\mu m^2$ single pixel) system with respect to CFD thresholds. Constant slew-rate and abence of Landau fluctuations in the 3D case result in a flat dependence.

To mitigate statistical biases and efficiency issues at the dynamic range limit, a high bandwidth readout system (> 3 GHz) is crucial. Such effects can be studied by leveraging the Poissonian distribution of radioactive decays. Using a two-object coincidence system, a DUT bias scan is performed requesting identical number of events at each step. The reference device (LGAD) is set to a known bias, yielding 100 % efficiency. An event frequency distribution is constructed at each step exploiting timestamp differences of consecutive events. The per bias rate is determined through Bayesian inference of a Poison fit, using the Gamma function as a conjugate prior [10]. The maximum achievable rate corresponds to the full efficiency point, assuming fixed source activity and geometrical acceptance. Relative efficiencies can subsequently be attributed for each bias, by comparing each point's corresponding rate with the one at the 100 % efficiency point.

Figure 3 displays the relative efficiencies for varying bandwidth limits and analog-to-digital (ADC) scaling as a function of bias voltage. A typical S-curve is observed in the 2 GHz - 10 mV series, achieving 100 % efficiency at ~ -36 V. When the analog scaling is doubled, without other system changes, there's a noticeable efficiency improvement, peaking at ~ -30 V. However, this configuration results in a 20 % reduction in the per point event rate, compared to the 10-mV scenario. The requirement of a minimum 3 ADC bin threshold in the Time-over-Threshold (ToT) for the trigger circuit to latch, alters the signal distribution by eliminating its faster component when scaling is increased. The charge and amplitude to ToT proportionality of the used transimpedance amplifier, biases the charge distribution towards higher values, even though the actual trigger threshold remains the same.

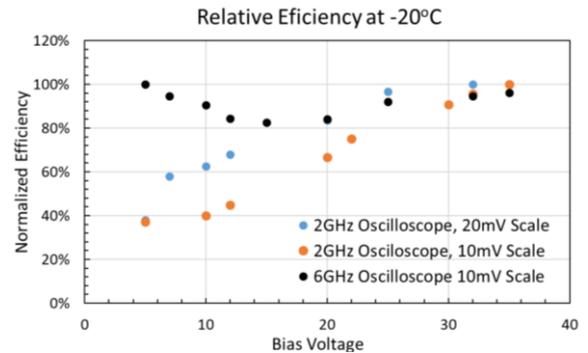

Fig. 3. Relative effieicny vs bias voltage for several ADC scaling values and badnwidth limits.

Increasing the ADC bandwidth from 2 to 6 GHz, while maintaining amplifier cut-off at 3 GHz, results to 100 % relative efficiency at the lowest bias, but decreases as voltage rises. Using a square windowing function, the frequncty of the highest harmonic is extracted from the Fast Fourier Transform (FFT) of the signal part of each waveform for biases of -5, -20, and -35 V (Figure 4). Bias noticeable affects signal composition, with the faster component dominating at higher fields. With decreasing bias, the signal fraction around 0.4 GHz increased from 20 % to 80 %. The convolution of a bandwidth cut, evolving signal population and normal efficiency S-curve expected as fields increase, account for the observed shape of the 6 GHz series in Figure 3. At lower biases, most signals fall below the bandwidth cut and are recorded. However, as the field intensifies, efficiency should increase, but part of the signal exceeds the bandwidth limit, reducing event rate. At the highest bias, 80 % of the signal exceeds the frequency cut-off, yet the increased efficiency compensates, recovering the rate of the slower component to levels observed at lower fields.

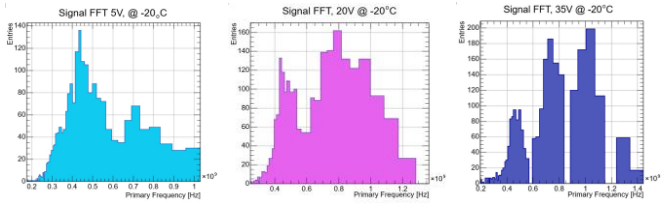

Fig. 4. Distributions of the frequency corresponding to the highest power harmonic for bias votlage of – 5 V, -20 V and -35 V at a 4000 events samnple. A compositon change is observed with respect to bias.

### 3. SPS pion Test beams

Properly addressing previously detailed issues and studying field non-uniformities require precise position resolution (~ 5 μm) combined with high statistics and high bandwidth electronics. To that end, an intensive 16-week long test beam campaign using 120 GeV pions was undertaken at CERN SPS. A timing telescope consisting of 6 DUTs and 2 Reference planes in coincidence was placed within a temperature controlled XPS cold box. Plane alignment is achieved through micrometric piezo-electric actuators, while the entire system is positioned between the forward and backward arm of a EUDET-based MIMOSA-26 [11] telescope. An FE-I4 planar module, attached to the back end of cold box, is used as an ROI trigger and alignment plane (Figure 5).

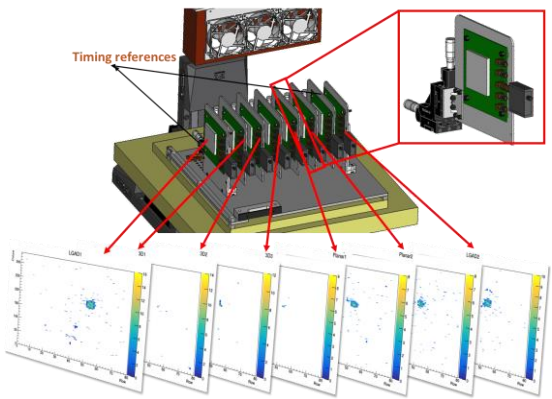

Fig. 5. CAD representation of the cold box with the DUT and timing reference planes. Occupancy distributions are used for plane alignemet.

System read-out is achieved through 2 synchronized 6 GHz oscilloscopes [12], while a fast SiGe-based first stage transimpedance amplifier [13] in conjunction with a 6 GHz second stage voltage amplifier (ZX-60V3 [14]) is attached to each plane. Data, generated in two streams (tracking, timing) are synchronized using the SPS master clock in conjunction with a trigger VETO, while readout is performed during the synchrotron acceleration cycles to eliminate dead time. A timing diagram of the architecture can be seen in Figure 6.

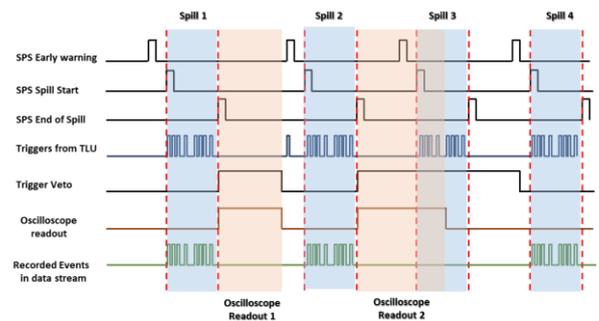

Fig. 6. Timing diagram of the trigger, readout and synchronisation system with respect to the VETO and SPS master clock siganls.

Individual event waveforms are analyzed using a multi-factor weighted approach to address increased noise, inherent to the high bandwidth and capacitance of the DUTs. A low pass filter, comprised of a rectangular transfer function followed by a Gaussian decrease, with a σ of 0.5 GHz and the -10 dB point centered at 2.4 GHz, is applied at the analysis stage to improve SNR and allow for signal smoothing (Figure 7, 8). Although for the reference sensor (LGAD) such an operation dramatically improves signal quality and lowers noise (Figure 7), for the 3D structure, it results in a 60 % degradation of amplitude and 20 % increase of calculated collected charge, without any advantages on SNR. The lower capacitance of the DUT (20 – 80 fF depending on geometry) compared to that of the reference (2 pF for 1 x 1 $mm^2$ LGADs) eliminates any potential gain from such an approach. In contrast, any low pass filtering will bias the signal statistics towards the high charge region, both due to the distorted signal shape and lower signal amplitudes, resulting to an enhanced rejection of the low amplitude tails, thus pushing the MPV of the Landau to a seemingly higher value.

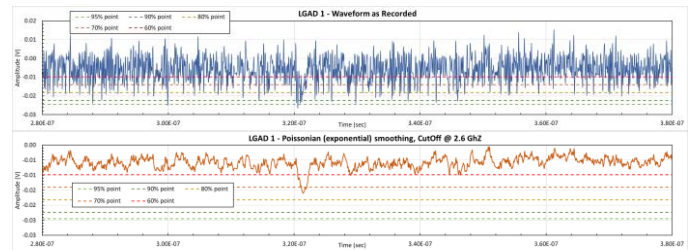

Fig. 7. As recorded and after a 2.4 GHz low-pass Gaussian filter LGAD signal. Dotted lines indicate percentages of the original signal amplitude.

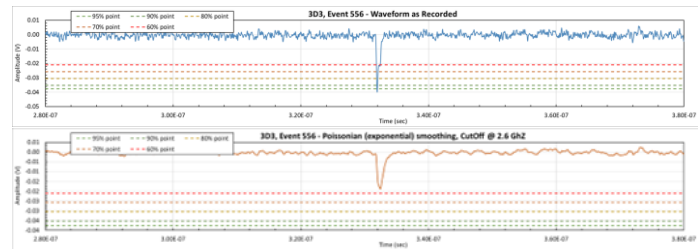

Fig. 8. As recorded and after a 2.4 *GHz* Gaussian filter 3D signal. Dotted lines indicate percentages of the original signal amplitude.

Using a Keizer window, the Fourier transform of the signal region of the Figure 7 & 9 event waveform is computed for the 3D and the LGAD sensors (Figure 9). While the unsmoothed 3D signal expands up to 5 GHz, the first harmonic of the LGAD signal debuts at 0.6 GHz. Appling the Gaussian low-pass filter with the 2.4 GHz cut-off frequency on the 3D signal, significantly reduces the higher frequency side of the spectrum. An appropriate readout electronics in this case would need to at least have a bandwidth of 5 GHz, whereas such an approach for an LGAD would be detrimental due to the higher induced noise without any gain on the signal side.

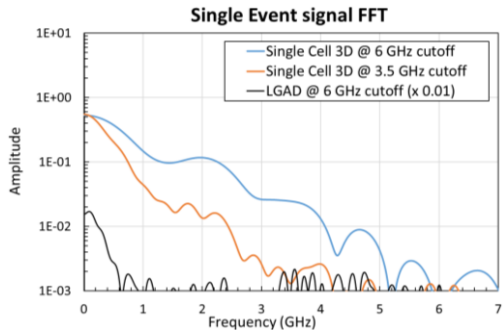

Fig. 9. Power Spectrum of an LGAD, 3D and Gaussian filtered 3D sigal. A 40 % power loss is observed for the smoothed 3D signal.

## 4. Fast front-end electronics

To adequately record high bandwidth signals, a discrete 16-channel read-out board based on a two-stage amplifier approach was designed (Figure 10). A two-stage single SiGe transistor amplifier geometry, with first stage configured as a transimpedance and the second as a voltage amplifier, was used. Provisions for individual channel shielding to further reduce noise are implemented, while a high frequency Rogers 3035 laminate is used in a five-layer PCB design with the signal plane encapsulated within isolated ground layers. Sensor bias is provided via a triaxial LEMO connector while keyed 8-fold coaxial high bandwidth mini-mcx connector arrays are placed unilaterally to the PCB.

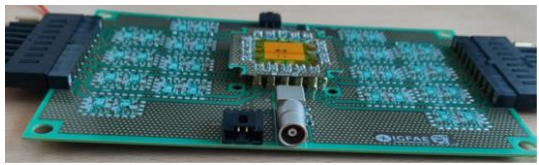

Fig. 10. Prototype of the hgih bandwidth versatile multi-channel PCB.

The DUT is placed on a passive 300 µm thick mezzanine board, populated with 18 coaxial mini-mcx connectors, to facilitate alignment and exchangeability. Peripheral components were optimized to ensure moderate gain fluctuation with frequency (Figure 11) through AWR simulations up to the 10 GHz limit. A mean noise (~RMS) of 1.2 mV for a gain of ~70 was achieved for the first iteration of the board, with an observed undershoot due to mismatch in lower frequencies.

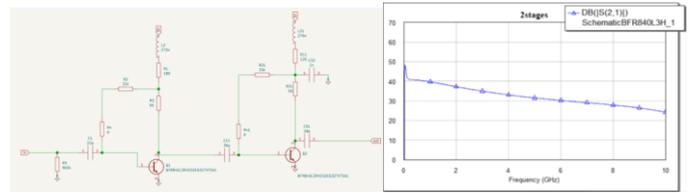

Fig. 11. Scematics of the circuit (left) and simulations (right) for up to 10 GHz of the amplification circuit of the two stage SiGe implementation.

## 5. Conclusion

In this paper, the bandwidth dependent charge and time resolution of high-speed 3D signals was discussed. Signals extend up to a limit of 5 GHz and adapted read-out electronics are necessary for an unbiased charge and noise measurement. Through an SPS-Pion test beam, time resolution studies are performed using a high bandwidth system, synchronized to the accelerator clock. The issue of higher induced noise in the larger capacitance LGAD device used as time reference is treated through a Gaussian low pass filter. Finally, a dedicated discrete electronics readout-out board with a ~10 GHz bandwidth limit was designed where DUTs are placed on a detachable mezzanine to enable rapid exchange and testing.